\documentclass{article}

\usepackage{arxiv}

\usepackage[utf8]{inputenc} %
\usepackage[T1]{fontenc}    %
\usepackage{url}            %
\usepackage{booktabs}       %
\usepackage{amsfonts}       %
\usepackage{nicefrac}       %
\usepackage{microtype}      %
\usepackage{lipsum}		%
\usepackage{graphicx} %
\graphicspath{ {./images/} }
\usepackage[authoryear]{natbib}
\usepackage{amsmath}
\usepackage{mathrsfs}
\usepackage{multirow}
\usepackage{bbm}
\usepackage{overpic,color}

\newcommand{\confcb}{
\begin{overpic}[width=\textwidth]{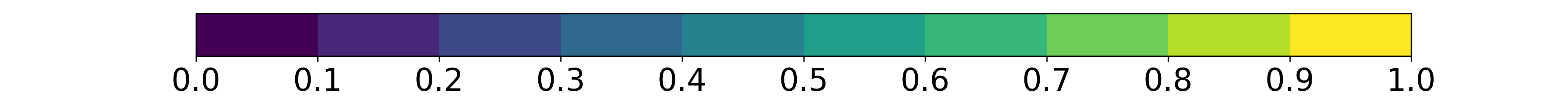}
       \end{overpic} 
}

\newcommand{\NAMEOFPICKER}{PickBlue}
\title{\NAMEOFPICKER: Seismic phase picking for ocean bottom seismometers with deep learning}

\author{Thomas Bornstein$^{1, 2, *}$, Dietrich Lange$^{1}$, Jannes Münchmeyer$^{2, 3}$, Jack Woollam$^{4}$,\\
\textbf{Andreas Rietbrock$^{4}$, Grace Barcheck$^{5}$, Ingo Grevemeyer$^{1}$, Frederik Tilmann$^{2, 6}$}\\

$^1$ GEOMAR Helmholtz Centre for Ocean Research Kiel, Wischhofstr. 1-3, 24148 Kiel, Germany\\
$^2$ GFZ, German Research Centre for Geosciences, Telegrafenberg, 14473, Potsdam, Germany\\
$^3$ Univ. Grenoble Alpes, Univ. Savoie Mont Blanc, CNRS, IRD, Univ. Gustave Eiffel, ISTerre, Grenoble, France\\
$^3$ Institute (GPI), Karlsruhe Institute of Technology, Karlsruhe, Germany\\
$^4$ Department of Earth and Atmospheric Sciences, Cornell University, 112 Hollister Drive, Ithaca, USA\\
$^6$ Institute for Geological Sciences, Freie Universität Berlin, Berlin, Germany\\
$*$ now at: Gempa GmbH, Heinrich-Mann-Allee 18/19, 14473 Potsdam, Germany
}

\date{}

\renewcommand{\headeright}{}
\renewcommand{\undertitle}{}
\renewcommand{\shorttitle}{\NAMEOFPICKER: Seismic phase picking for ocean bottom seismometers with deep learning}

\begin{document}
\maketitle

\begin{keypoints}
 \item We assembled a database of Ocean Bottom Seismometer waveforms and manual P and S picks, on which we train PickBlue, a deep learning picker.
 \item Our picker significantly outperforms pickers trained with land-based data with confidence values reflecting the likelihood of outlier picks.
 \item The picker and database are available in the SeisBench platform, allowing easy and direct application to OBS traces and hydrophone records.
\end{keypoints}
~\\

\begin{abstract}
Detecting phase arrivals and pinpointing the arrival times of seismic phases in seismograms is crucial for many seismological analysis workflows. For land station data machine learning methods have already found widespread adoption.
However, deep learning approaches are not yet commonly applied to ocean bottom data due to a lack of appropriate training data and models.
Here, we compiled an extensive and labeled ocean bottom seismometer dataset from  15  deployments in different tectonic settings, comprising  $\sim$90,000 P and $\sim$63,000 S manual picks from 13,190 events and 355 stations.
We propose \NAMEOFPICKER, an adaptation ot the two popular deep learning networks EQTransformer and PhaseNet.
\NAMEOFPICKER~joint processes three seismometer recordings in conjunction with a hydrophone component and is trained with the waveforms in the new database. The performance is enhanced by employing transfer learning, where initial weights are derived from models trained  with land earthquake data.
\NAMEOFPICKER~significantly outperforms neural networks trained with land stations and models trained without hydrophone data.
The model achieves a mean absolute deviation (MAD) of 0.05\,s for P waves and 0.12\,s for S waves.
We integrate our dataset and trained models into SeisBench to enable an easy and direct application in future deployments.
\end{abstract}
~\\

\begin{plsummary}
Ocean bottom seismometers are seismic stations on the seafloor. Just like their counterparts on land, they record many earthquakes on three component sensors but are additionally equipped with underwater hydrophones. 
To determine the location of an earthquake, seismologists must precisely measure the arrival times of seismic waves.
This task can be and is traditionally carried out by human analysts. Machine learning has been highly successful in taking over this job for seismic records from onshore. However, the noise and the signal are different in the ocean environment. For example, the recordings can contain whale songs and water layer reverberations and are disturbed by ocean bottom currents. These differences have hindered the direct application of existing machine learning pickers for ocean bottom data. We have assembled an extensive database of ocean bottom recordings and trained artificial neural networks to use the underwater hydrophone information and cope with the ocean noise environment. We demonstrate that the resulting machine learning picker picks are similar to those of human experts. We make the database and machine learning picker available with a standard interface so that it is easy for other scientists to apply them in their studies.
\end{plsummary}

\section{Introduction}

Determining the arrival times of seismic phases in seismograms is crucial for detecting and locating  earthquakes. Accurate and precise onset times of P and S phases are a precondition for many seismological applications such as magnitude estimation, focal mechanisms determination, or travel time tomography. 
Consequently, automated onset picking of earthquake arrivals has been an active field of research for several decades \cite{allen-automatic-1982}. 
Until the recent advance of machine learning (ML) pickers, most phase picking algorithms were picking P arrivals \cite{baer_picker_1987,lomax12,sleeman-robust-P-picking-1999}, while S phase picking algorithms were less common \cite{diehl-automatic-S-picking-2009,sleeman-single-2003-P-S-picker} due to the increased complexity of phase determination in the coda of the P wave, and the generally lower frequency content of the S wave.
They often require a prior estimate of the hypocentre in order to rotate horizontal components and to identify the approximate time window where S is expected.

Most automated phase picking algorithms are optimized for land stations 
\cite{sleeman-robust-P-picking-1999,diehl-P-Picker-2009,leonard_multi-component_1999,kuperkoch_automated_2010}. 
However, as many tectonically interesting regions are submarine, accurate phase picking is required for ocean bottom seismometers (OBS) as well.
Unfortunately, picking phase onsets of OBS waveforms is more difficult: confounding factors, such as water column reverberations, distant seismic or volcanic signals, sounds of marine mammals and anthropogenic noise from shipping, generally lead to lower quality phase arrivals than on land.
In addition, the free-fall deployment procedure leads to tilt noise and unknown orientations of the horizontal channels \cite{crawford_compliance_1998}.
This makes polarisation techniques, commonly employed for S-onset determination \cite{diehl-automatic-S-picking-2009}, more difficult to apply.
On the upside, OBS data often include an additional hydrophone channel not available at onshore seismometers. 
The hydrophone channel, in combination with the vertical channel, can help to distinguish arrivals traveling through the water column (e.g., water multiples or direct waves from marine mammal soundings) from those traveling through the solid earth. 

Most of the time, algorithms developed for land stations are applied to marine seismological data with little modification, thereby foregoing any benefit from hydrophone data. This was true for conventional phase picking algorithms \cite{lieser_splay_2014,kuna_mode_BLANCO_2019} some years ago but is still the case for modern ML algorithms \cite{wu-submarine-catalogue-2022}. 
In particular, \citet{ruppert_alaska_2023} applied the deep-learning model EQTransformer with the original land-station trained weights to the full amphibious AACSE dataset, pointing out that the original EQTransformer does not necessarily generalize well to the AACSE OBS. 

Here, we train and test different variants of the \NAMEOFPICKER~ phase picker targeting OBS instruments using deep learning. Following common usage in the ML community, we refer to these as phase picking models.
Our models, trained directly on OBS data, are able to learn the characteristics of OBS three-component data and also factor in the hydrophone information.

We base our picker on two recent neural network architectures designed for picking, EQTransformer~\cite{mousavi_EQTRANSFORMER_2020} and PhaseNet~\cite{zhu2019phasenet}.
We extend both network designs by adding an additional hydrophone component as a fourth input channel, using 
 the implementation integrated within SeisBench \cite{Woollam-etal-SEISBENCH-2022}, an open-source toolbox and model repository for ML in seismology.
To train the OBS picking models, we compiled an extensive database of annotated local event waveforms recorded with free-fall OBS, including magnitudes, locations, and manually picked arrival times.
We demonstrate the superior performance of these picking models compared to the equivalent pickers trained without hydrophone traces. Finally, we integrated our models and the trained models weights into SeisBench to enable the straightforward application to new data.

\section{Data}

\begin{figure}
    \centering
    \includegraphics[width=\textwidth]{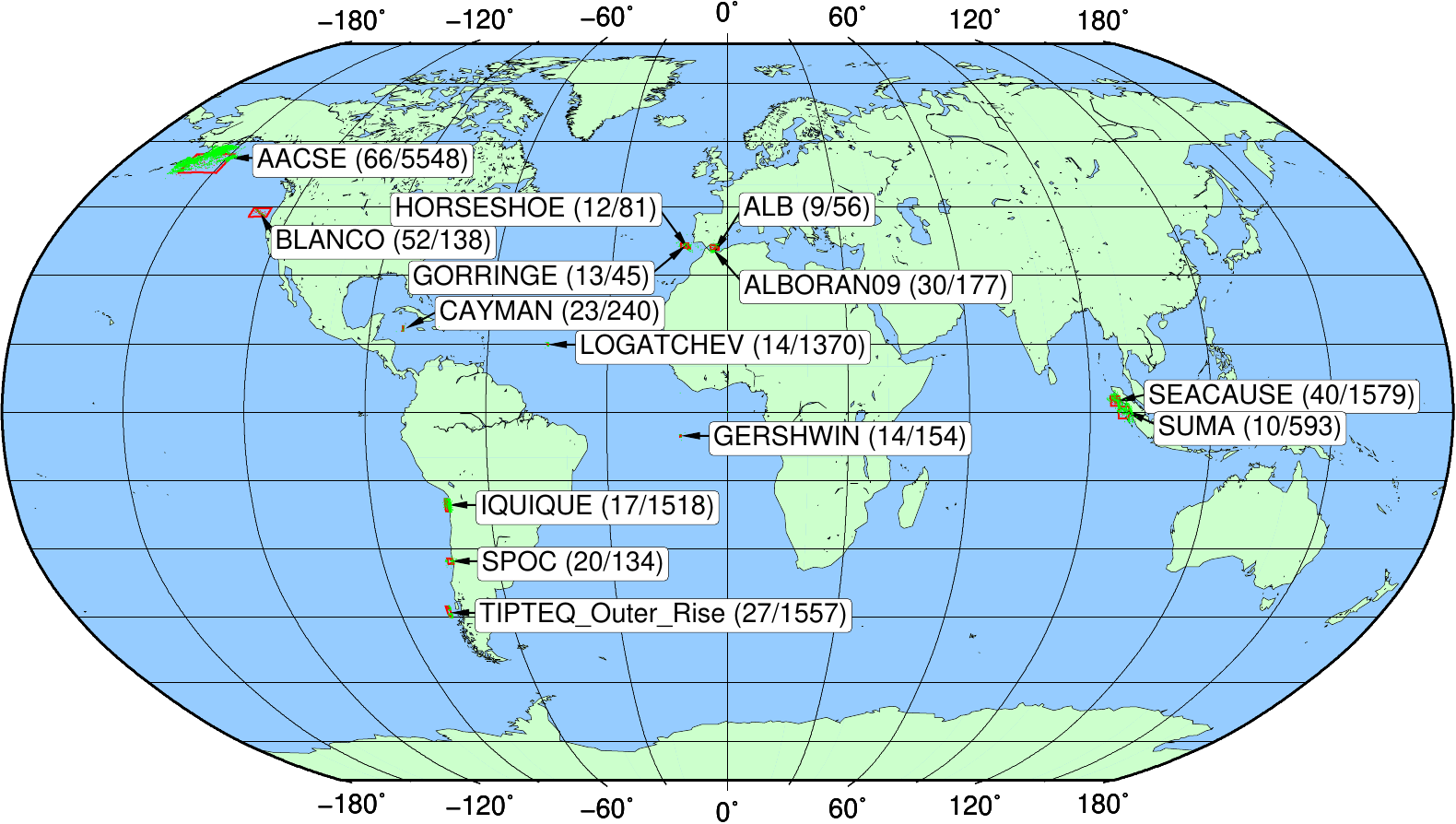}
    \caption{Global map showing the distribution of the OBS networks used for training the \NAMEOFPICKER\ picking algorithms. Labels indicate the dataset name and numbers of OBS stations and events used. Red boxes encircle the OBS networks. Stations and events are shown with red and green dots, respectively. ALA is not shown as the station set is identical to that of AACSE.  %
    }
    \label{fig:data_global}
\end{figure}

We compiled an extensive database of waveforms from local earthquakes in various submarine tectonic environments, mostly four-component data, but also some OBH data, i.e., with the hydrophone as the only channel.
With each waveform, we provide  manually labeled P and/or S phase picks and---for most deployments---also station locations and estimated earthquake locations and magnitudes.
Fig.~\ref{fig:data_global} shows the global distribution of the OBS deployments included and the recorded seismicity. Table~\ref{tab:exp_times_count} provides additional information, including references for the contributed deployments.\\

The complete OBS dataset contains manually picked phases from 15 deployments and a total of 355 stations (Table~\ref{tab:exp_times_count}). The dataset comprises 13,190 events, 109,210 traces and 153,338 picks (about 90,000 P and 63,000 S picks). The data represent a variety of tectonic environments, seismometer and hydrophone types. The data include 38,419 4-component waveforms; 35,654 waveforms have only seismometer data (no hydrophone), and 8,187 traces are for hydrophone-only instruments (Table~\ref{tab:traces_comps}). For the remainder of the traces, 1 or 2 seismometer components are missing due to equipment malfunction. Magnitudes range from 0.1 to 5.8 (Fig.~\ref{fig:source_magnitudes}).

We split each deployment 
into a training set (65\%), a development set (10\%) and a holdout test set (25\%). When we added the AACSE subset, we kept its pre-defined split ratio (70\%/15\%/15\%). The effective split ratio for the whole dataset was 66.8\%/12.8\%/20.4\%. 
We randomly distributed entire events over the splits, i.e., ensure that all traces belonging to an event are part of the same split, thereby avoiding knowledge leaks from intra-event waveform similarity.

The OBS data is provided through SeisBench \cite{Woollam-etal-SEISBENCH-2022} and can be used for other ML learning tasks. No explicit noise traces are included but noise samples can be generated by extracting the waveform ahead of the P arrival. 

\begin{table}
    \centering
    \caption{Table with OBS deployments used for training the \NAMEOFPICKER\ networks}
    \scriptsize
        \centering
    	\begin{tabular}{lccllrrrp{0.22\textwidth}}
        	\hline Experiment$^a$ 
        	& Tectonic       & OBS      & Start date & End date & \#wave-  & \#P & \#S & Reference  \\
        	& setting$^b$      & type$^c$ &            &          & forms        &     &     & \\
        	\hline \\
        	AACSE &  S & BB & 2018-05-12 & 2019-08-31 & 52,645 & 40,051 & 39,725    & \citet{ruppert_alaska_2023},\citet{barcheck_aacse_dataset_2023}, XO (2018-2019)\\
        	ALA$^d$ &  S &  BB& 2018-10-01 & 2019-02-09 & 315   & 302 & 93   & \citet{barcheck_alaska_2020}, XO (2018-2019)\\
        	ALB & I/CR &   SP& 2016-09-15 & 2016-12-03 & 274   & 201 & 192          & I.  Grevemeyer, pers. comm.\\
        	ALBORAN2009 &I/CR &SP & 2009-08-13 & 2010-01-16 & 2,252 & 1,622 & 1,900  & \citet{grevemeyer_seismicity_Alboran_2015}\\
        	BLANCO      &T&BB& 2012-09-26 & 2013-09-23 & 2,882  & 2,850 & 961  & \citet{kuna_mode_BLANCO_2019},  X9 (2012-2013)\\
        	CAYMAN & R &SP & 2015-04-03 & 2015-04-16 & 2,302    & 1,582 & 1,665    & \citet{grevemeyer_constraining_utraslow_2019}\\
        	GERSHWIN & R & SP & 2000-05-03 & 2000-05-12 & 834     & 811 & 28   & \citet{tilmann04} \\
        	GORRINGE & I & SP & 2013-10-11 & 2014-03-25 & 404    & 343 & 178    & \citet{grevemeyer_seismotectonics_Horseshoe_2017}  \\
        	HORSESHOE &I & SP& 2012-04-15 & 2012-10-14 & 696  & 677 & 175     & \citet{grevemeyer_seismotectonics_Horseshoe_2017}\\
        	IQUIQUE & S & SP & 2014-12-09 & 2016-10-29 & 6,913   & 6,712 & 1,608    & \citet{petersen_relationship_2021} \\
        	LOGATCHEV &R & SP & 2009-01-18 & 2009-03-26 & 11,427 & 9,428 & 6,571  & \citet{grevemeyer_microseismicity_Logatchev_2013}\\
        	SEACAUSE & S & SP & 2005-10-16 & 2006-03-02 & 16,177  & 15,103 & 4,104    & \citet{tilmann10} \\
        	SPOC & S & SP & 2001-10-13 & 2001-12-01 & 1,339     & 1,332 & 62     & \citet{thierer05} \\
        	SUMA & S  & SP/BB  & 2008-06-06 & 2009-02-09 & 2,468   & 2,345 & 954        & \citet{lange10} \\
        	TIPTEQ & OR& SP & 2004-12-12 & 2005-01-27 & 8,280     & 6,734 & 5025    & \citet{tilmann_seismicity_2008}   \\[0.5\baselineskip]
        	\hline
        	total &&& 2000-05-03 & 2019-02-09 & 109,208  &   90093 & 63241  &  \\ \hline
     \multicolumn{9}{l}{$^{a}$ The OBS datasets are accessible through the SeisBench framework.  The waveforms and metadata comprise $\sim$35\,GB.}\\
     \multicolumn{9}{l}{$^{b}$  I=Intraplate, R=Ridge, S=Subduction Zone, T=Transform Fault,  OR=Outer Rise.}\\
    \multicolumn{9}{l}{$^{c}$ BB=Broadband OBS, SP=Short Period OBS.}\\
    \multicolumn{9}{p{\textwidth}}{$^{d}$ The ALA picks are based on the identical station set as the much larger AACSE dataset. A small fraction of picks (272 picks, or 0.18\% of the complete dataset) were also independently picked for ALA, by a different human analyst. Since the overlap is very small this will only have a negligible influence on the performance evaluation.}
   	\end{tabular}
    \label{tab:exp_times_count}
\end{table}

\section{Methods}

Several deep-learning models for seismic phase picking have recently been published
\cite{ross_GPD_2018,woollam_BasicPhaseAE_2019,zhu2019phasenet,mousavi_EQTRANSFORMER_2020,soto2021deepphasepick}.
We base \NAMEOFPICKER~on these previously published models.
For the application to OBS data, we focus on PhaseNet and EQTransformer.
We chose these models because of their favorable performance identified in the recent benchmark study of \citet{benchmarkpaper}. 
For ease of reading, in the following, we refer to our adapted versions of the models as BluePhaseNet and BlueEQTransformer, and to those other versions that we also studied simply as PhaseNet and EQTransformer.

Our training and evaluation procedure follows the steps outlined in \citet{benchmarkpaper}.
The neural networks are trained with manually picked phase arrival times of known earthquake waveforms and noise samples taken from the same dataset. 
We train the networks to reproduce a characteristic function where Gaussian peaks with amplitude 1 and half-width 0.2~s are centered on the manual picks of P and S arrivals, respectively, identically to the procedure in the benchmarking study.

For evaluation, the ML picker is provided with a longer waveform segment (30 or 60 s, depending on the picker architecture) which contains the manual pick at a random position. Based on this waveform sample, two characteristic functions are calculated, which can be thought of as (non-calibrated) measures of the probability of a sample to be a P-wave or S-wave, respectively. We then examine a 10\,s window from the output that contains the manual pick at a random location within the window.  We evaluate both the phase type classification, P or S, and the accuracy of the onset determination, with a focus on the latter.
The onset time is given by the time of the peak value of the characteristic function for the respective phase.
The target is to be as close as possible to the manual prediction. Of course, particularly for low signal-to-noise ratios or otherwise ambiguous scenarios, the manual pick might be inexact. Therefore, it is not to be expected that a deep-learning picker can perfectly reproduce the manual pick times. Nonetheless, the average difference between manual and deep learning picks is a solid indicator of the model performance.

\begin{figure}
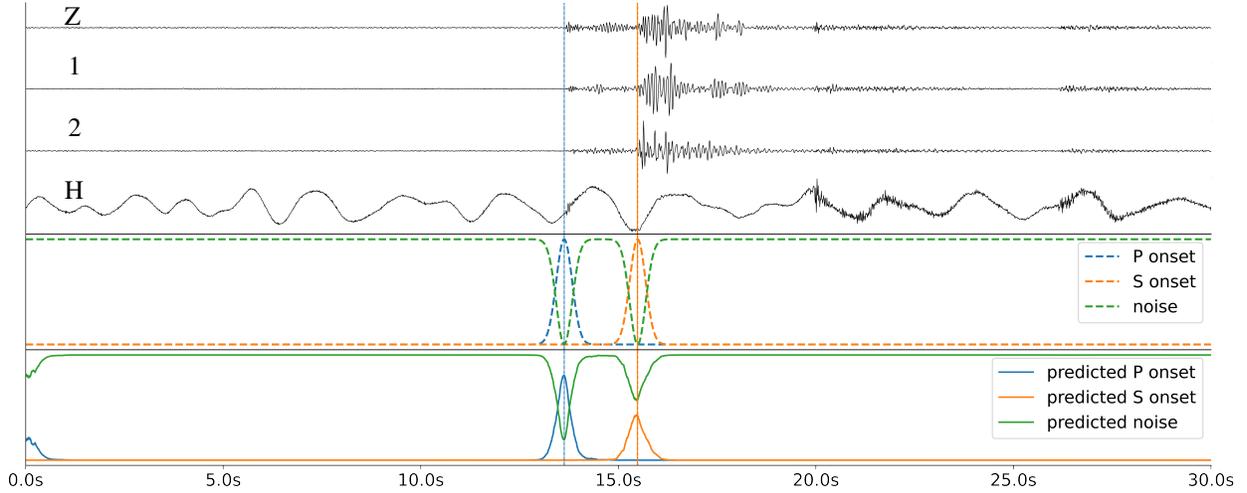

    \centering
     \begin{overpic}[width=\textwidth]{images/ex_trace_pred_pn_47-01.png}
     \setlength{\fboxsep}{3pt}
           \put(5.2,38.0){Z}
           \put(5.5,34.0){1}
           \put(5.5,29.0){2}
           \put(5.2,24.0){H}
    \end{overpic}
    \caption{Example of an input waveform sample for BluePhaseNet. From top to bottom:  trace with four components after preprocessing (resampling, cutting of waveform, normalization); ground truth characteristic functions (P, S phase arrivals and noise) used for training the models; characteristic function predicted by BluePhaseNet prediction. Z: vertical component; 1 and 2:  two horizontal components; H:  Hydrophone channel.}
    \label{fig:ex_trace}
\end{figure}

\subsection{Models}

We trained two models for picking the OBS data. Both models have originally been trained on land station data.

\begin{description}
    \item [Earthquake Transformer (EQTransformer):]
        EQTransformer is a model for joint event detection, phase detection, and onset picking 
        \cite{mousavi_EQTRANSFORMER_2020}. 
        EQTransformer uses a stack of convolutional layers~\cite{lecun1995convolutional}, long short-term memory cells (LSTMs)~\cite{hochreiter1997long} and self-attention layers~\cite{luong2015effective,yang2016hierarchical}. It consists of a down-sampling section of CNNs and max-pooling layers, followed by an encoder using residual CNNs and LSTMs. Then, self-attention layers add contextual information enabling the model to focus on the most important parts of the sequence. Subsequently, three separate decoders map the information to three probability sequences (detection, P phase, S phase). EQTransformer consists of 378,928 trainable parameters. It expects 60\,s long input traces for all channels. 
    \item [PhaseNet]:
        PhaseNet \cite{zhu2019phasenet} is a U-Net \cite{ronneberger2015u}, consisting of a convolutional and a deconvolutional branch.
        During the down-sampling process, four convolutional stages condense the information. 
        Then, in the up-sampling process of deconvolutions, the model expands and converts this information into probability distributions~\cite{zhu2019phasenet}.
        PhaseNet consists of 268,499 trainable parameters, about 30 percent less than EQTransformer.
\end{description}

\subsection{Training workflow}

As discussed above, the dataset contains traces with missing components due to data gaps, too large timing offset between the components and different hydrophone and seismometer types (Table \ref{tab:traces_comps}).
For these traces, missing data and components are replaced with zeros.
This ensures that the model input always contains four channels and at the same time, makes the models robust to missing data in future application scenarios.

The traces are resampled to 100\,Hz.
We ensure that our data selection leads to a uniform distribution of pick locations within the input windows.
Where traces are too short, missing samples were padded with zeros.  
We normalize the amplitude by dividing each component by its maximum amplitude independently.
By normalizing each component individually, we avoid either hydrophone or velocity (seismometer) amplitudes being reduced to values close to zero, as the data were not corrected to physical units and absolute amplitudes depend on digitizer gain settings. 
Last, we subtract the mean and the linear trend for all channels. 
Fig.~\ref{fig:ex_trace} shows an example input after the preprocessing steps used for PhaseNet. 

The original EQTransformer makes intensive use of augmentations during training. 
Therefore, we also apply the following augmentations to the EQTransformer training process to each sample, with probability $p$: 
(i) adding a secondary earthquake signal into the empty part of the trace ($p=0.3$), 
(ii) adding a random level of Gaussian noise ($p=0.5$), 
(iii) randomly inserting gaps ($p=0.2$),
(iv) dropping one or two channels ($p=0.3$). 
We do not apply the additional random wrap-around shifting of the traces, as employed in the original implementation. This is not required, as the lengths of the original waveforms (at least 60~s noise before the P pick is available for 7\% of the traces and for 36\% in case of S picks) ensures that the pick can naturally occur at any location in the trace (see also Fig.~\ref{fig:trace_lengths}).

 While we input the three seismometer components unfiltered to the networks, we high-pass filter (0.5\,Hz) the hydrophone data before feeding it to the network as a fourth component.
Filtering proves beneficial due to strong low-frequency noise on the hydrophone traces resulting  from the broadband characteristics of hydrophones. 

We experiment with two strategies for model initialization.
    (1) standard training: initializing the models with random weights and training them on the OBS dataset tailored to the OBS domain;
    (2) transfer learning: initializing the models with weights from the same models but pre-trained on STEAD and INSTANCE datasets, which both consist of land station data only; the hydrophone related channels are still initialized with random weights. 
Transfer learning takes advantage of the model performance in a similar domain, thus resulting in faster convergence and often higher generalizability, particularly where the number of training data samples used for the pre-training significantly exceeds that available for the specialized training \cite{pan2009survey,TransferLearningJozinovic,TEAM}.

For cross-domain applications, the datasets used for training or pre-training must be of the same distance range as the later application domain \cite{benchmarkpaper}. Furthermore, source models trained on large datasets generally yield the best performance after fine-tuning. As the OBS data contains local earthquakes, we choose the PhaseNet and EQTransformer models trained with INSTANCE and STEAD datasets for initializing the weights.

We tested different hyperparameters and determined the following set as best suited: batch size 2048 (EQTransformer)/ 1024 (PhaseNet); learning rate 0.001/0.01; epochs 200/400. We use the Adam optimizer~\cite{kingma_ADAM_2014}.
The models with minimum loss of the development subset were found after 185 (EQT) and 240 epochs (PhaseNet), respectively. Those models were subsequently used in the evaluation. 
The training of the models took about 19 hours on an NVIDIA A100-40 GPU with 40\,GB memory. 

\section{Results}

In the following, we compare the performance of the different models.
In order to retain clarity of presentation, we start with a discussion of the preferred models, i.e., the models showing the best overall performance.
These are the EQTransformer and PhaseNet models using transfer learning with pre-training on INSTANCE, which we refer to as BlueEQTransformer and BluePhaseNet, respectively.
However, the PhaseNet model trained without pre-trained weights achieves effectively the same performance.
We then provide a more detailed analysis and comparisons to other model variants.
Unless specified otherwise, all analyses were conducted for the preferred models.

\subsection{Onset time determination}

We use the residuals, i.e., the differences between the ML pick and the onsets picked by human analysts (as stored in the dataset metadata), to evaluate the quality of onset time picking. We analyze the modified root mean squared error (RMSE), the mean absolute deviation (MAD), and the modified mean absolute error (MAE).
Specifically, we calculate the RMSE and MAE by (arbitrarily) defining outliers as residuals outside the interval $\pm$1.0\,s, and taking them into account with a value of 1~s. In this way, we avoid overly strong influence of the outliers on the metrics.
We separately report the outlier fraction according to this definition.

\begin{figure}
\centering
       \begin{overpic}[width=\textwidth]{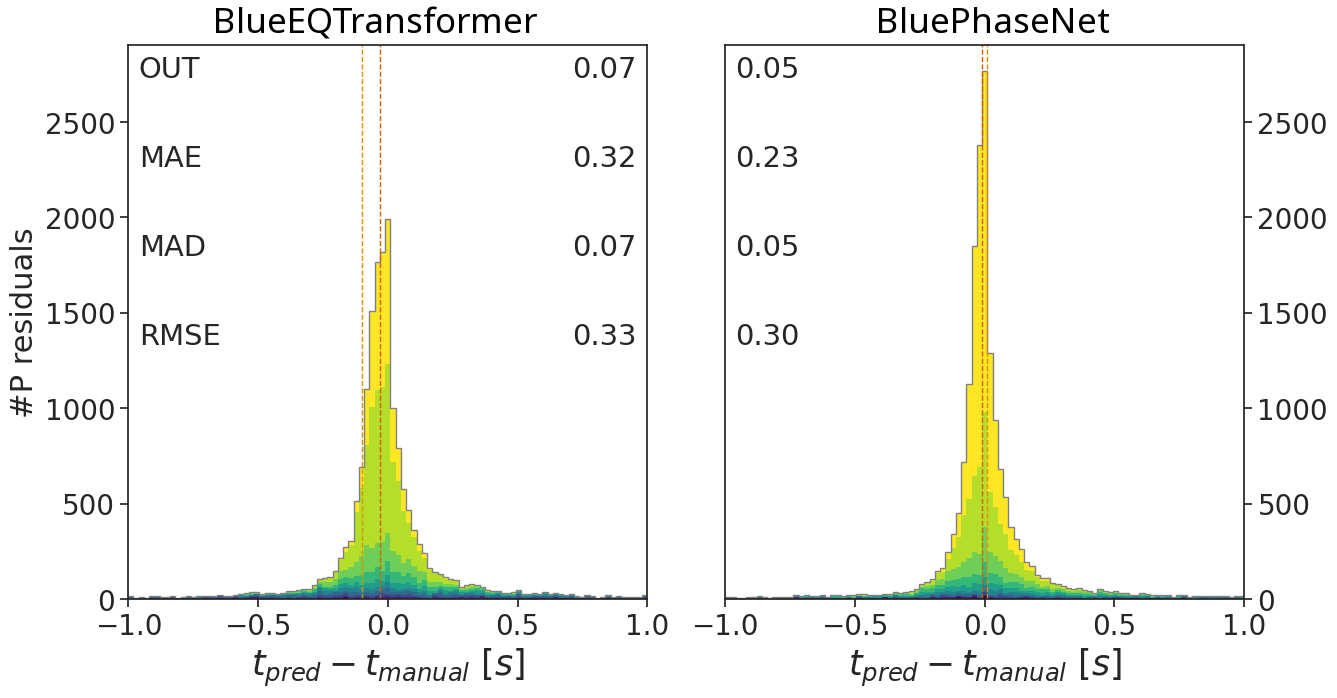}
                \put(12,18){\framebox{\mbox{\LARGE{a}}}}
                \put(56,19){\framebox{\mbox{\LARGE{b}}}}

       \end{overpic}
    \\
       \begin{overpic}[width=\textwidth]{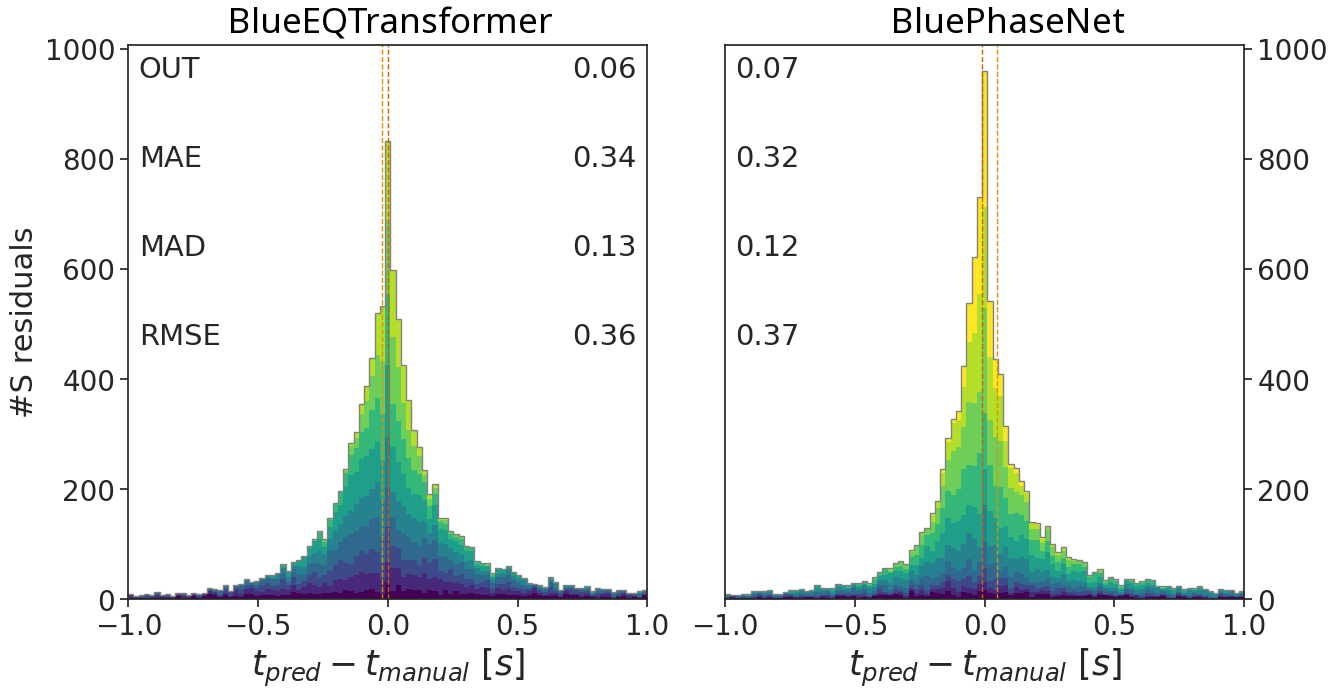}
                \put(12,19){\framebox{\mbox{\LARGE{c}}}}
                \put(56,19){\framebox{\mbox{\LARGE{d}}}}
       \end{overpic}
     \confcb       
     \caption{Histogramm of residuals between the manual picks and picks by BlueEQTransformer and BluePhaseNet (pre-trained on INSTANCE) for P phases (panels a,c) and  S phases (panels b,d); the bin width is 0.02~s. The  vertical dashed lines mark the median (orange) and mean (yellow) residual but note that the mean is strongly influenced by outliers. The histogram columns are subdivided by color-coded pick confidence (peak of the characteristic function for P or S arrivals); each segment length corresponds to the frequency of residuals with the respective confidence. OUT: Fraction of outliers (residuals with absolute value $>$1.0 s, i.e., outside the window shown), MAE: Mean absolute error, MAD: Median absolute deviation. RMSE: Root mean square error. Note that the y-axis differs for the lower and upper panels. Note that in the calculation of MAE and RMSE, residuals with absolute values exceeding 1\,s were set to $\pm$1\,s in order to reduce the dependence of these measures on outliers. See Fig.~\ref{fig:tl-inst-out} for a view of the same distribution but extending to $\pm10\,s$.}
     \label{fig:performance-EQT-PHnet}
\end{figure}

Fig.~\ref{fig:performance-EQT-PHnet} shows the residual distribution of P and S wave picks using the preferred BlueEQTransformer and BluePhaseNet models.
For P onsets, BluePhaseNet outperforms BlueEQTransformer with MAEs of 0.23~s vs. 0.32~s and RMSEs of 0.30~s vs. 0.33~s, whereas both models show comparable results for S onsets.
The median is very close to 0.0~s for both models, showing that there is virtually no bias with respect to the manual picks.
The central distribution resembles a Laplacian.
For P picks, 90\% of picking errors are smaller than 0.62~s and 0.46~s for BlueEQTransformer and BluePhaseNet;
95\% are within 1.33~s/0.99~s.
For S picks, 90\% are below 0.67~s/0.71~s, 95\% below 1.17~s/1.2~s.

Comparing P residuals to S residuals, we confirm that the determination of S arrivals is more difficult, as can be seen, for example, in the nearly double MAD for S compared to P.
We attribute the lower performance for S picks to two main factors.
First, S arrivals usually have lower signal-to-noise ratios due to the typically much higher noise levels of OBS horizontal components. Also, overlap with the P coda and precursors from a basement or Moho Sp conversion can create ambiguous onsets.
Furthermore, our training dataset contained substantially fewer examples of S arrivals, giving the models less possibility to learn the characteristics of S arrivals.
Nonetheless, the performance for S arrivals is still excellent, in particular considering that traditional pickers are usually unable to detect S arrivals on OBS recordings at all.

\begin{figure}
  \begin{overpic}[width=1.0\textwidth]{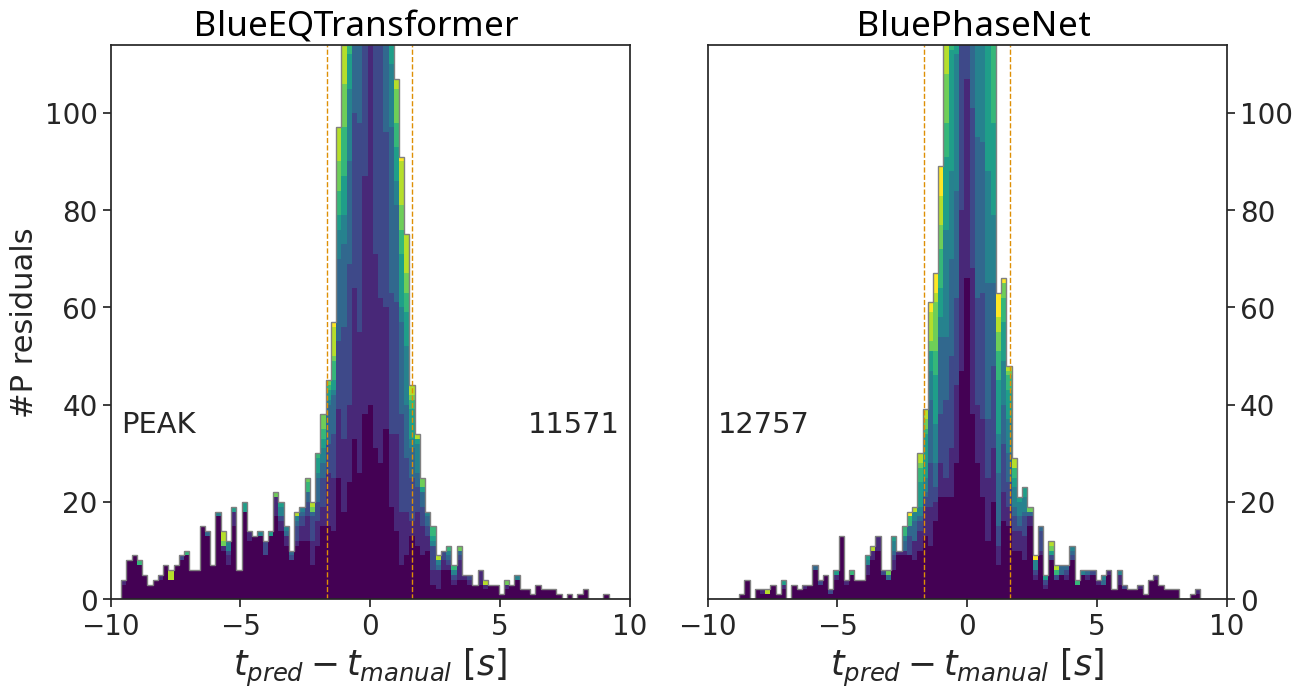}
                \put(10,45){\framebox{\mbox{\LARGE{a}}}}
                \put(56,45){\framebox{\mbox{\LARGE{b}}}}
       \end{overpic}    \\
       \begin{overpic}[width=\textwidth]{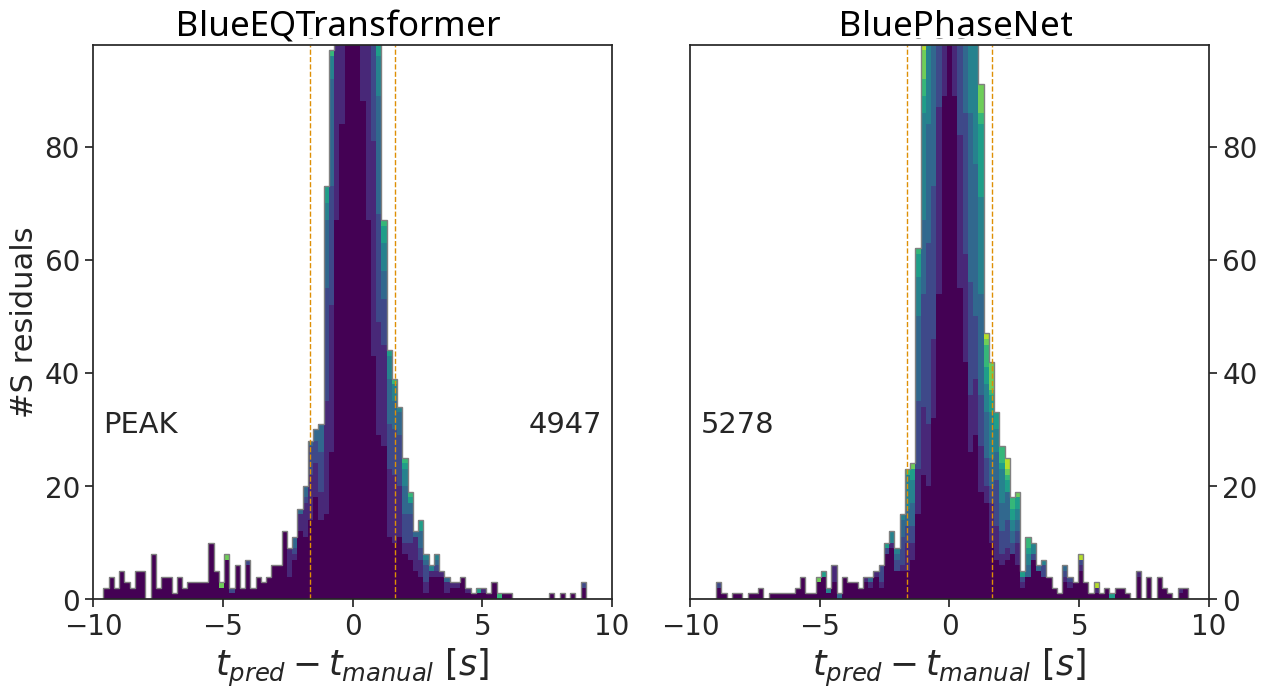}
                \put(10,45){\framebox{\LARGE{{c}}}}
                \put(56,45){\framebox{\mbox{\LARGE{d}}}}
       \end{overpic}
  \confcb       
  \caption{Histogram of residuals between the manual picks (as Fig. \ref{fig:performance-EQT-PHnet} but  showing the full range of possible residuals from -10 to +10\,s and with a bin width of 0.2~s) and picks by BlueEQTransformer and BluePhaseNet (pre-trained on INSTANCE)  for P phases (panels a,c); and  S phases (panels b,d). The y-axis range is chosen to emphasize the tails of the distribution, with PEAK showing the peak value of the histogram, which will be near zero on the $x$ axis, but be far outside the $y$ range shown. 90\% confidence intervals are marked with yellow vertical lines.}
  \label{fig:tl-inst-out}
\end{figure}

Fig.~\ref{fig:tl-inst-out} shows a zoomed out version of the error distribution, focussing on the outliers.
Up to about $\pm 2.5$~s, the residuals follow a Laplacian distribution.
Outside this interval, they follow a more uniform or low-sloped triangular distribution.
We will refer to these picks as blunders in line with previous usage for describing analyst picks \cite{nmsop_is11_4}.
Although both BlueEQTransformer and BluePhaseNet show blunders, there are some subtle differences. For BlueEQTransformer, the central part (i.e. up to $\pm2.5 s$) appears fairly symmetric, but blunders earlier than the manual pick occur more often than blunders in the coda of the P wave.
This results in a mean value shifted to early picks (-0.07\,s).
Conversely, blunders appear fairly symmetric with respect to the manual pick for BluePhaseNet, whereas, regarding S onset residuals, the central part is very slightly skewed towards later arrivals, resulting in too late average picks (mean 0.05\,s).

\begin{figure}
\centering
  \includegraphics[width=0.8\textwidth]{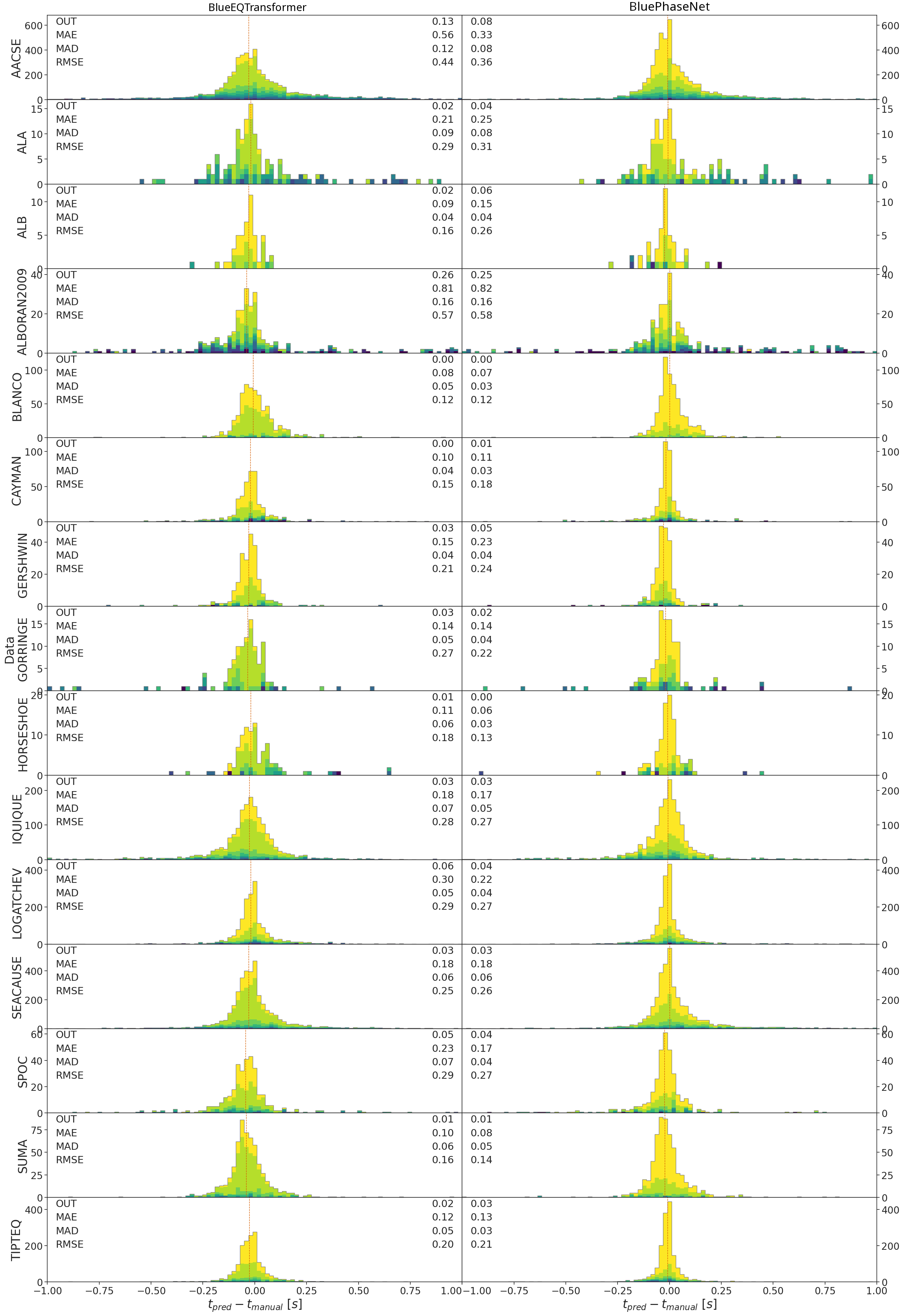}
  \confcb
  \caption{P residuals for the individual experiments between -1.0 and \texttt{+}1.0\,s. BlueEQTransformer (BluePhaseNet), pre-trained on INSTANCE, are shown in the left (right) column, respectively. The panels are labeled to the right with the name of the dataset. Fig.~\protect\ref{fig:tl-inst-p-exp-out} shows the same data with the range extended to $\pm$10\,s. }
  \label{fig:tl-inst-p-exp}
\end{figure}

\begin{figure}
\centering
   \noindent\includegraphics[width=0.8\textwidth]{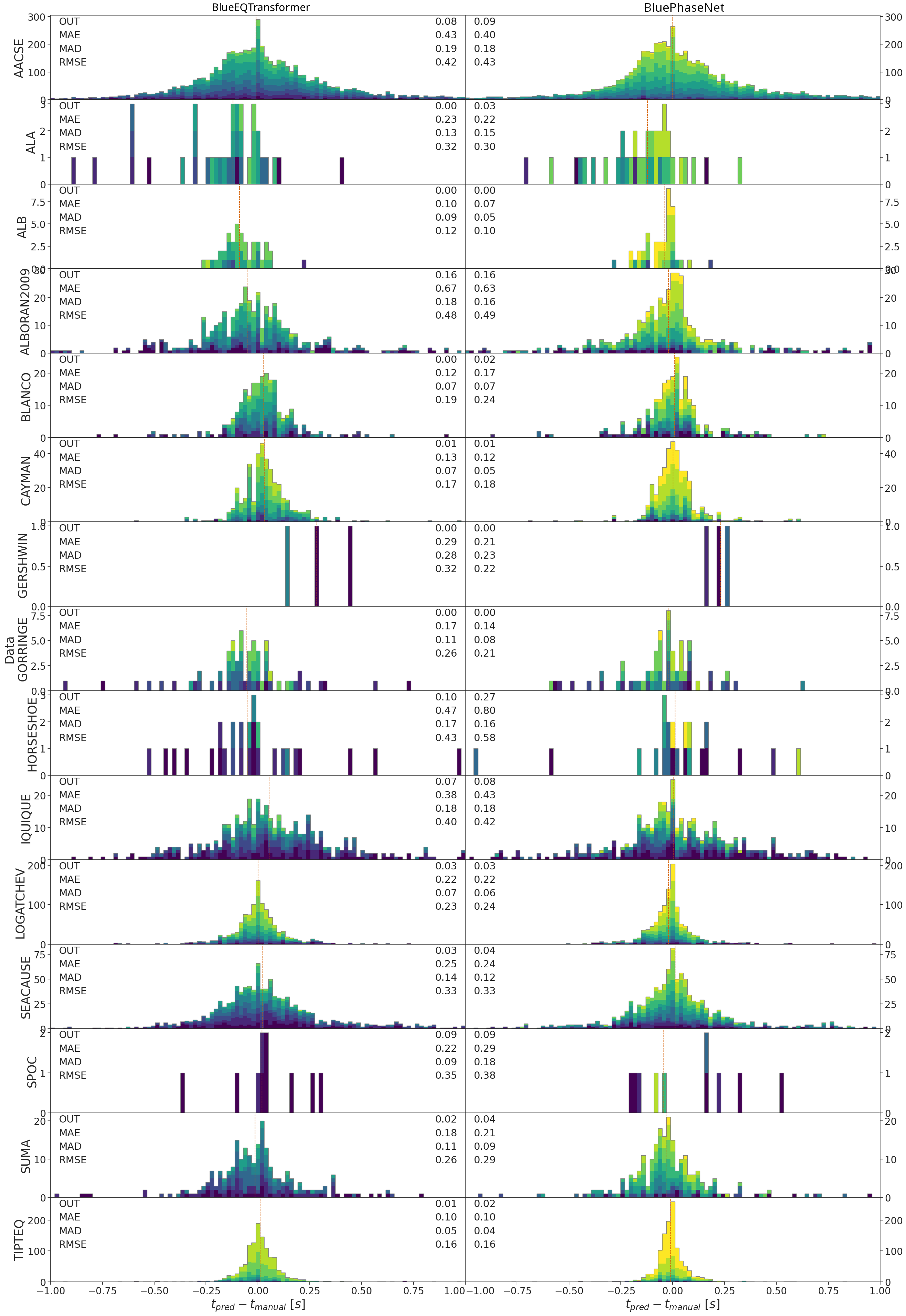}
   \confcb
  \caption{As Fig.~\ref{fig:tl-inst-p-exp} but for S residuals. Fig.~\protect\ref{fig:tl-inst-s-exp-out} shows the same data with the range extended to $\pm$10\,s.}
  \label{fig:tl-inst-s-exp}
\end{figure}

Fig.~\ref{fig:tl-inst-p-exp} and \ref{fig:tl-inst-s-exp} break down the performance by the individual deployments.
The overall performance varies strongly across datasets, likely due to differences in noise environment,  magnitude distribution, network geometry and station types (broadband or short period).
Notably, the slight superiority of BluePhaseNet relative to BlueEQTransformer is valid across most datasets.
On datasets where BlueEQTransformer outperforms BluePhaseNet, the performance difference is minor.
Nonetheless, both pickers are viable across many different settings and return consistently small errors.
The great majority of (BlueEQTransfomer/BluePhaseNet) picks have errors below 0.2~s: 77~\%/81~\% for P, 63\%/65~\% for S for the whole dataset; the ranges for individual deployments are 53-96\%/56-98\% for P, 36-94~\% 33-93\%/52-95\% for S. 
While, for each deployment, every second S pick of BluePhaseNet has an absolute error below 0.2~s, BlueEQTransformer achieves the same only with every third pick.

\subsection{Classification of phase types}

As mislabelled phase types are challenging for association algorithms and result in large residuals during earthquake location, we investigated the reliability of the P vs S classification. %
For the comparison, we used 14,404 traces from the test set which have any combination of P and S picks. We classify a predicted P pick as  misclassified if it has higher confidence than the predicted S pick on the same trace and if it is closer to the manual S pick than to the manual P pick; vice versa for predicted S picks. Picks with confidence below 0.1 were ignored.
In general, phase identification was very accurate (see Table~\ref{tab:my_label} for the misclassification matrix). 
Only 11 P picks (0.25\%) were wrongly classified as S arrivals from BluePhaseNet. In turn, 20 (0.4\%) S phases were wrongly classified as P phases. 
In sum, BlueEQTransformer has even fewer misclassification. Only 16 P picks (0.32\%) were classified as S arrival, and 6 S picks (0.14\%) were classified as P arrivals. 
Figures~\ref{fig:ex-pn-s-near-p}-\ref{fig:ex-eqt-p-near-s} show examples of misclassified events for P and S phases for both BluePhaseNet and BlueEQTransformer.

\subsection{Confidence and picking quality}

Both models provide time series of probabilities for P and S phase onsets.
We interpret the peak values of these time series as the confidence of the model in the pick.
Here, we try to understand how this value relates to the accuracy of the pick with respect to the manual pick.
In general, we expect smaller residuals for higher pick confidence.
This way, confidence values could serve as proxies for pick quality. 
Depending on the use case, a picker that refuses to return a pick for a few cases might be preferable over a picker that returns more, but low quality picks.

Each of the bins in Fig.~\ref{fig:performance-EQT-PHnet}-\ref{fig:tl-inst-s-exp} consists of a sorted stack of confidences, giving an impression of how the confidences map to picking errors. The length of each segment corresponds to the frequency of residuals with the respective confidence.
Three major observations can be made. 
First, both models pick P waves with higher confidence than S waves.
This becomes even more obvious when considering individual experiments (Fig.~\ref{fig:tl-inst-p-exp} vs. Fig.~\ref{fig:tl-inst-s-exp}).
Second, even though there is a considerable difference between the experiments, high confidence value cluster around 0 residual, i.e., the higher the confidence the lower the expected residual.
Third, picks with low confidence still follow an approximately Laplacian distribution centered near 0.0 but with higher uncertainties (Fig.~\ref{fig:performance-EQT-PHnet}).
Crucially, almost all of the outliers have low confidence (Figure \ref{fig:tl-inst-out}).

\begin{figure}
   \begin{overpic}[width=\textwidth]{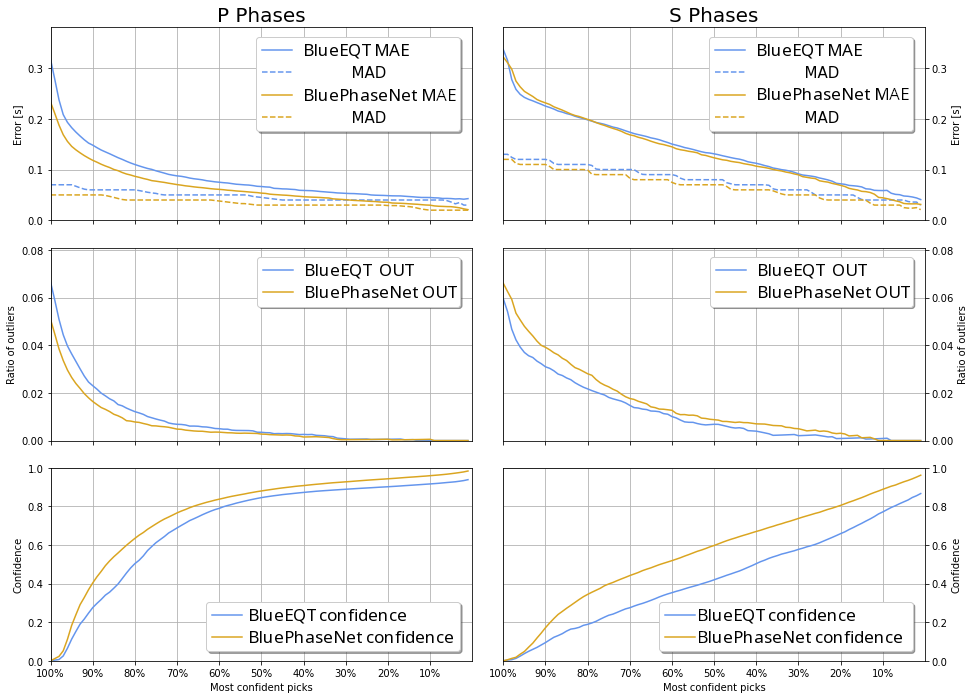}
    \put(08,63){    \colorbox{white} { \framebox{\mbox{\LARGE{a}}}}  }  
    \put(54,63){    \colorbox{white} { \framebox{\mbox{\LARGE{b}}}} }

    \put(08,41){    \colorbox{white} { \framebox{\mbox{\LARGE{c}}}}  }
    \put(54,41){    \colorbox{white} { \framebox{\mbox{\LARGE{d}}}} }

    \put(08,19){    \colorbox{white} { \framebox{\mbox{\LARGE{e}}}}  }
    \put(54,19){    \colorbox{white} { \framebox{\mbox{\LARGE{f}}}}  }
    \end{overpic}
  \caption{Dependency of MAE, MAD (panels a,b), and number of outliers (panels c,d) for different subsets of the dataset, sorted by confidence value, such that the value on the left side of each plot corresponds to the full dataset (as shown in Fig.~\ref{fig:performance-EQT-PHnet}), whereas subsequently stricter confidence thresholds are applied to select only a set fraction of the whole dataset corresponding to the most confident picks. (e,f) shows percentiles of confidence value. Model training based on INSTANCE pre-training.}
  \label{fig:most_conf}
\end{figure}

Figure~\ref{fig:most_conf} explores the relationship between confidence and picking errors in a systematic manner.
We plot MAE, MAD and the outlier fraction, depending on which fraction of the most confident picks are considered. 
For example, the MAE  drops significantly for both P and S picks if the least confident 10\% are omitted. For S picks, we observe a knee in MAE and OUT curves at about 5\%. Further reduction leads to a further but more gradual reduction in MAE. 
On the other hand, the MAD for P waves decreases only very marginally when the least confident picks are omitted, whereas a strong reduction is seen in the number of outliers.
The drop in MAE is thus almost exclusively driven by a reduction in the number of outliers.
As the MAD is a robust measure of the spread of the central peak, this implies that there is hardly any dependence of the confidence value on the pick quality as long as the correct phase has been identified;  low confidences instead indicate a much higher chance of having misidentified another waveform feature, e.g., a later arrival or a local maximum in the noise time series as a phase, and thus produced an outlier.

For P arrivals, the number of outliers decreases until only 60\% of the picks are retained and stays almost constant afterward (Figure~\ref{fig:most_conf}a,c).
For S arrivals, no such clear cutoff can be identified, with outlier fractions decreasing steadily until only 30\% to 40\% of the picks are retained.
P and S waves also differ in the distribution of confidence values.
While S confidences are distributed almost uniformly, the P confidences tend towards higher values, e.g., the 60\% most confidence picks already have confidence values around 0.8.
These observations should be taken into account when selecting confidence thresholds in applications.
They suggest that P confidence values at higher confidences might be less reliable than their S counterparts; this is also reflected in the more continuous degradation in performance for S arrivals compared to their P counterparts.
Interestingly, confidence curves are almost identical across the two different models.

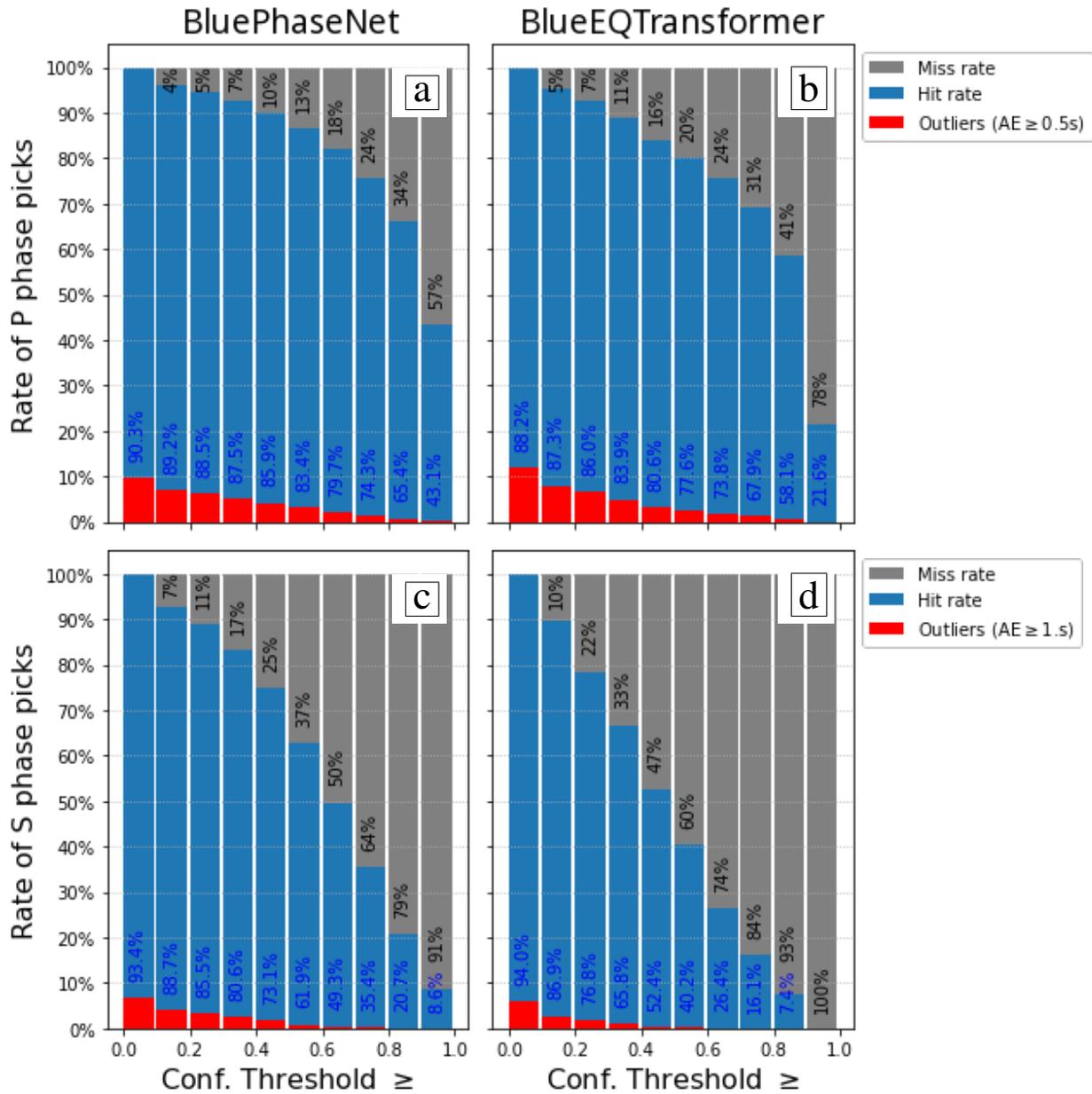
\begin{figure}
  \begin{overpic}[width=\textwidth]{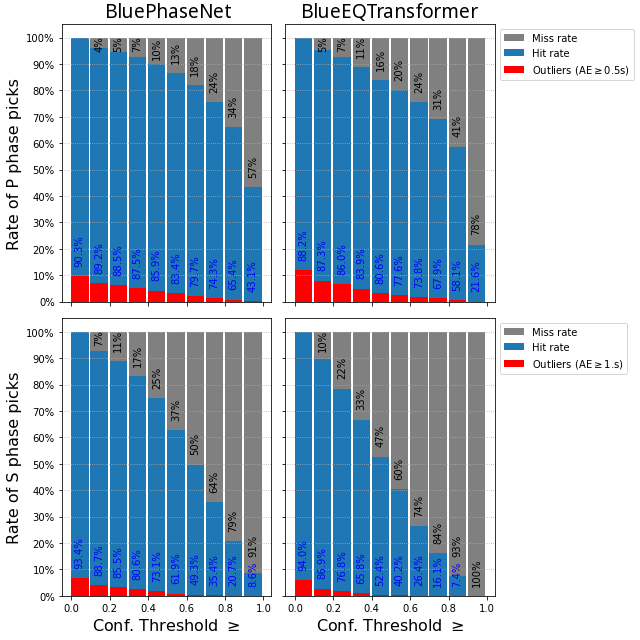}
    \put(35,91){    \colorbox{white} {  \framebox{\mbox{\LARGE{a}}}}              }
    \put(70,91){    \colorbox{white} { \framebox{\mbox{\LARGE{b}}}}              }
    \put(35,45){    \colorbox{white} { \framebox{\mbox{\LARGE{c}}}}              }
    \put(70,45){    \colorbox{white} {  \framebox{\mbox{\LARGE{d}}}}              }
  \end{overpic}
  \caption{Hit rate, miss rate, and number of outliers for different confidence thresholds, for models with INSTANCE pre-training. Black numbers specify the number of misses whereas blue numbers mark the number of hits that are not considered outliers. Note the different absolute error thresholds for picks to be considered outliers, noted in the legend.}
  \label{fig:outl_thres}
\end{figure}

Following the finding that residual error statistics are largely driven by outlier fractions with only minor variations of residuals in the central block, the choice of threshold in an application should primarily be guided by the relation of outliers, hit rate and miss rate.
Fig.~\ref{fig:outl_thres} visualizes the tradeoff of these parameters for different thresholds for both models.
For this analysis, we set a tighter threshold for P pick outliers (0.5\,s) to account for the typical accuracy expected for P picks and to make the outlier fraction more visible in the plot. 
Due to the more steady decrease in hit rate for S arrivals than for P arrivals, in general, a lower threshold is advisable for S arrivals than for P arrivals.
The exact choice of parameters depends on the dataset and the planned downstream analysis.

\section{Discussion}

\subsection{Effectiveness of transfer learning}

For all results presented so far, we used the preferred models, BluePhaseNet and BlueEQTransformer.
These models were pre-trained on INSTANCE and then fine-tuned on the OBS data.
In the following, we discuss the impact of this transfer learning and compare the results to transfer learning on STEAD.

\begin{figure}
  \begin{overpic}[width=\textwidth]{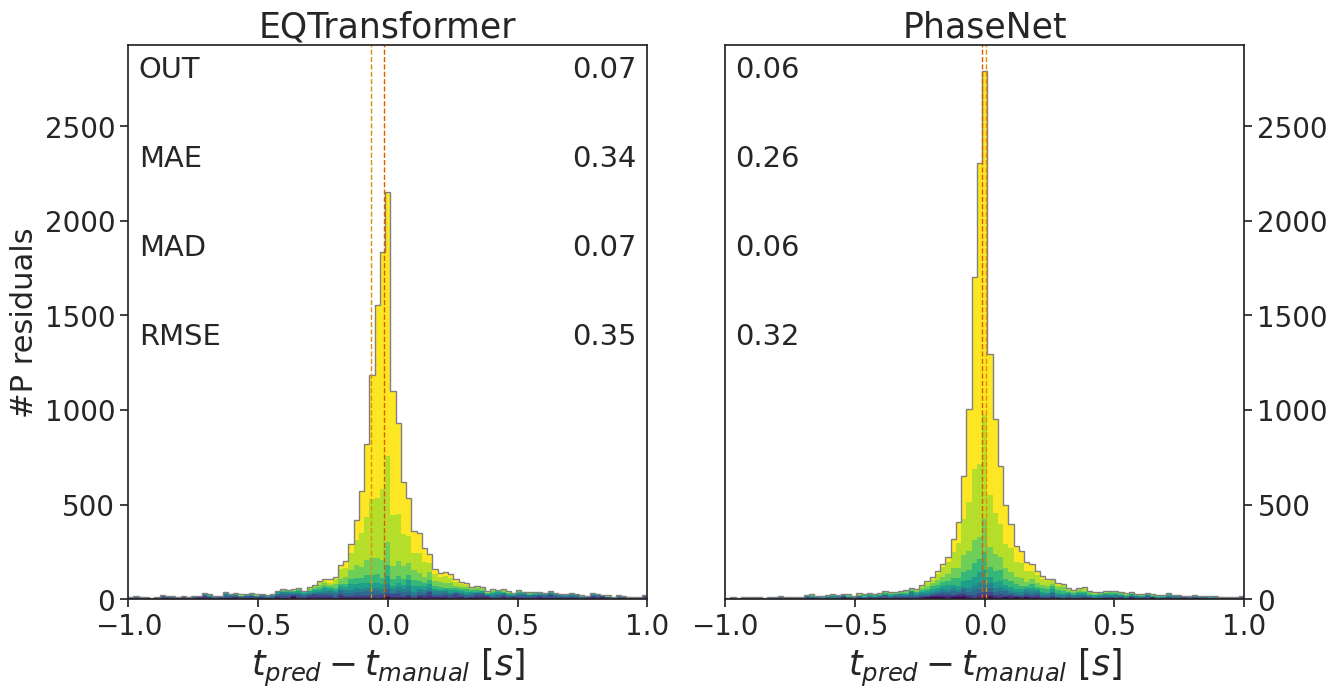}
                \put(12,10){\framebox{\mbox{\LARGE{a}}}}
                \put(56,10){\framebox{\mbox{\LARGE{b}}}}
       \end{overpic}    \\
  \begin{overpic}[width=\textwidth]{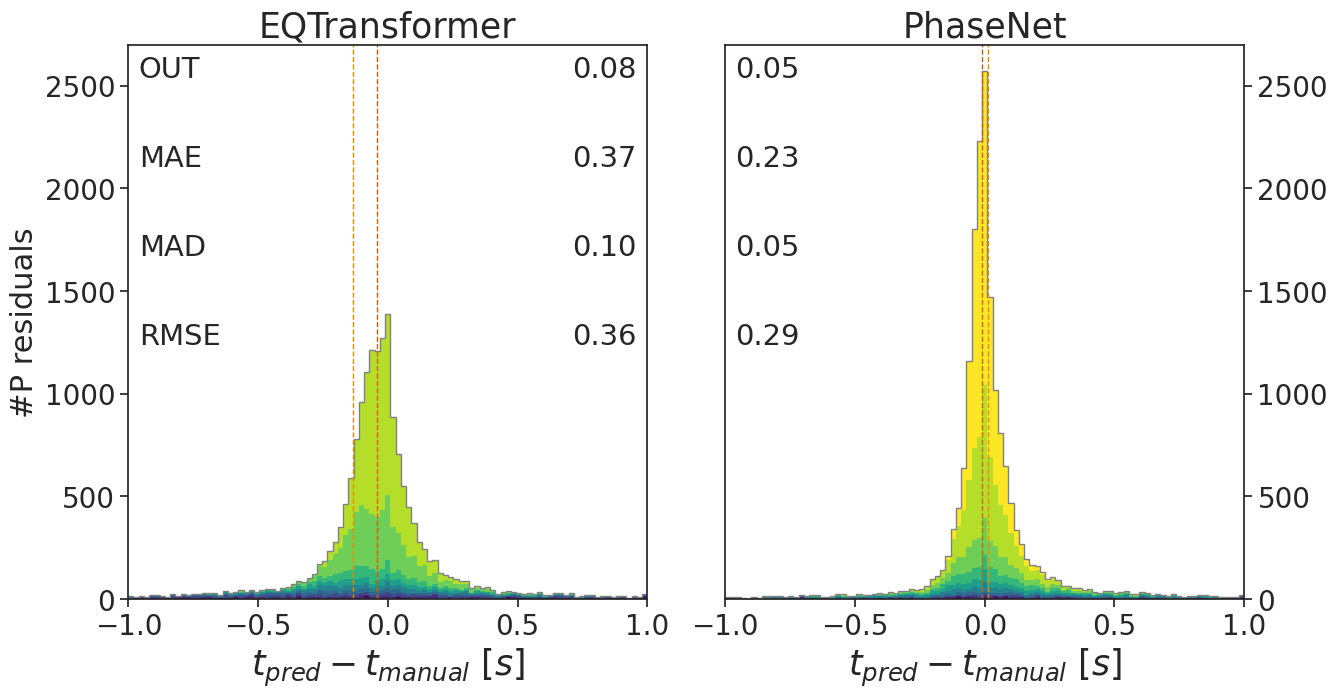}
                \put(12,10){\framebox{\mbox{\LARGE{c}}}}
                \put(54,10){\framebox{\mbox{\LARGE{d}}}}
       \end{overpic}    
\confcb       
\caption{P residuals with confidence stacks for different pre-training set-ups. (a,b) Models pre-trained on STEAD, then trained on OBS dataset. (c,d) Models trained on OBS dataset without pre-training. The full range to $\pm$10 s is shown in Fig.~\ref{fig:performance-p-pretrain-variants-out} and the equivalent plots for S residuals are shown in Fig.~\ref{fig:tl-stead-s} and~\ref{fig:obs-pure-s} for pre-training with STEAD and without pre-training, respectively. }
\label{fig:performance-p-pretrain-variants}
\end{figure}

Fig.~\ref{fig:performance-p-pretrain-variants} shows the performance in terms of P wave residuals of models without transfer learning and with transfer learning from STEAD.
EQTransformer shows substantially worse performance without transfer learning whereas PhaseNet is not only easy to train with randomly initialized weights but also does not show any benefit from transfer learning. 
We hypothesize that the impact of transfer learning on EQTransformer is larger than on PhaseNet due to the deeper and more complex architecture.
In particular, PhaseNet might benefit from the implicit parameter sharing across positions through the fully convolutional architecture and from the residual connections.
In addition, the EQTransformer without transfer learning generally exposes lower confidence than the one using transfer learning.
These patterns are similar for S waves.

Regarding the appropriate source dataset, we observe differences between EQTransformer and PhaseNet.
For EQTransformer, differences between pre-training on STEAD or INSTANCE are small and in both cases, pre-training improves performance.
For PhaseNet, pre-training on INSTANCE leads to better performance than on STEAD. In fact, the performance of the STEAD pre-trained PhaseNet is even worse than that of the non-transfer learned PhaseNet.
We hypothesize that the diversity of data in STEAD is insufficient and, therefore, leads to an overly adapted version of PhaseNet.
However, further experiments are required to validate this hypothesis.

Comparing the results on the individual deployments (Figs.~\ref{fig:tl-inst-p-exp}, \ref{fig:tl-stead-p-exp}, \ref{fig:tl-stead-s-exp}, and \ref{fig:obs-pure-p-exp}), there is substantial variability about the best performing model and training scheme.
While it is difficult to derive clear insights due to the diverse behavior, for some datasets a certain training scheme, i.e., pre-training either STEAD or INSTANCE or no pre-training at all, seems to be the best option irrespective of the model.

In conclusion, the analysis for individual data sets suggests 
that transfer learning often only yields moderate benefits and that its benefits depend on the targeted data set.
The already considerable number of training examples in the dataset we assembled, might explain the measurable but overall limited benefits of transfer learning.

\subsection{Significance of the hydrophone component and training on OBS data}

A key difference of OBS recordings to typical land-based recordings is the existence of an additional hydrophone component.
Our models incorporate this component as an additional input channel.
In addition, the noise environment in the oceans is different from that on land, as described in the Introduction. In this section, we compare the performance of \NAMEOFPICKER\ to that of the equivalent pickers trained on land data, and to three-component EQTransformer and PhaseNet networks trained on the OBS dataset.  

\begin{figure}

    \centering
  \begin{overpic}[width=0.9\textwidth]{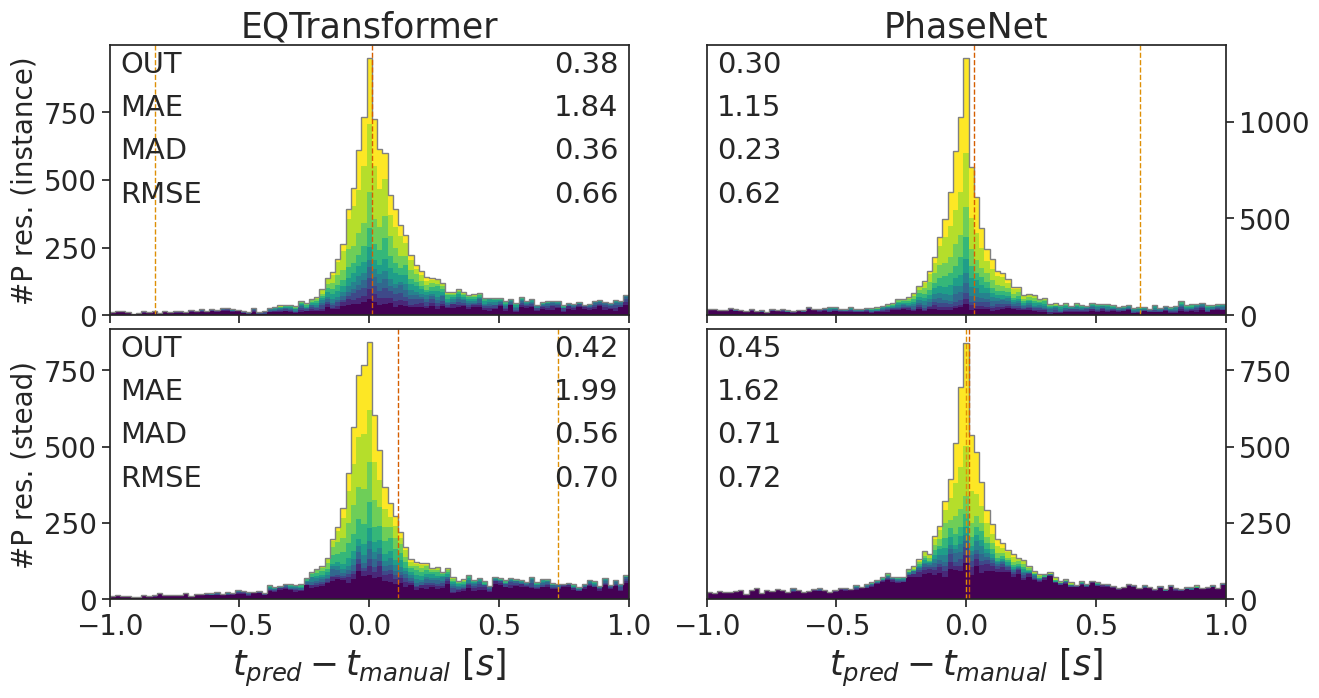}
                 \put(10,32){\framebox{\mbox{\LARGE{a}}}}
                 \put(54,32){\framebox{\mbox{\LARGE{b}}}}
                 \put(10,10){\framebox{\mbox{\LARGE{c}}}}
                 \put(54,10){\framebox{\mbox{\LARGE{d}}}}    
       \end{overpic}    

  \begin{overpic}[width=0.9\textwidth]{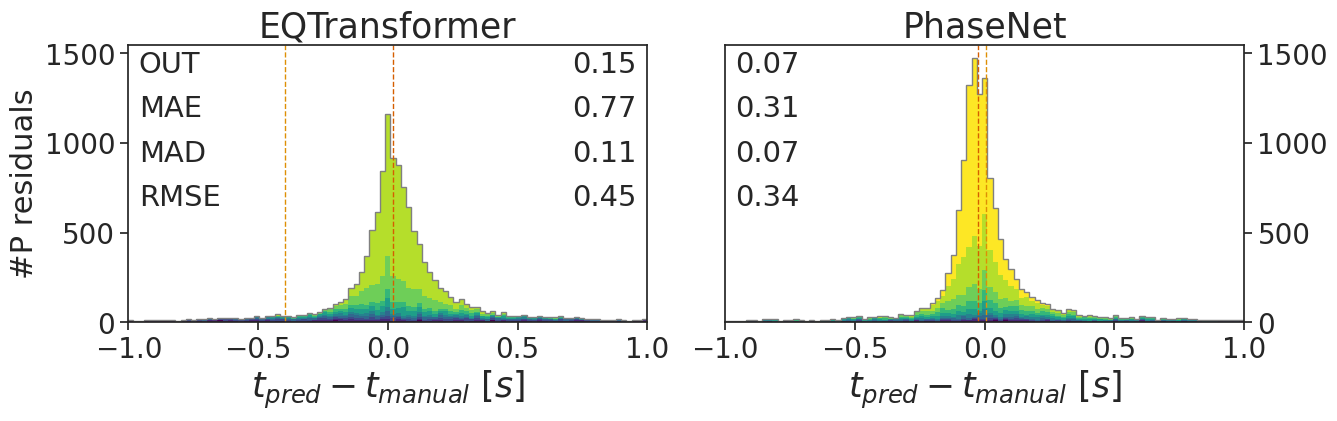}
                \put(10,10){\framebox{\mbox{\LARGE{e}}}}
                \put(56,10){\framebox{\mbox{\LARGE{f}}}}
       \end{overpic}    

  \begin{overpic}[width=0.9\textwidth]{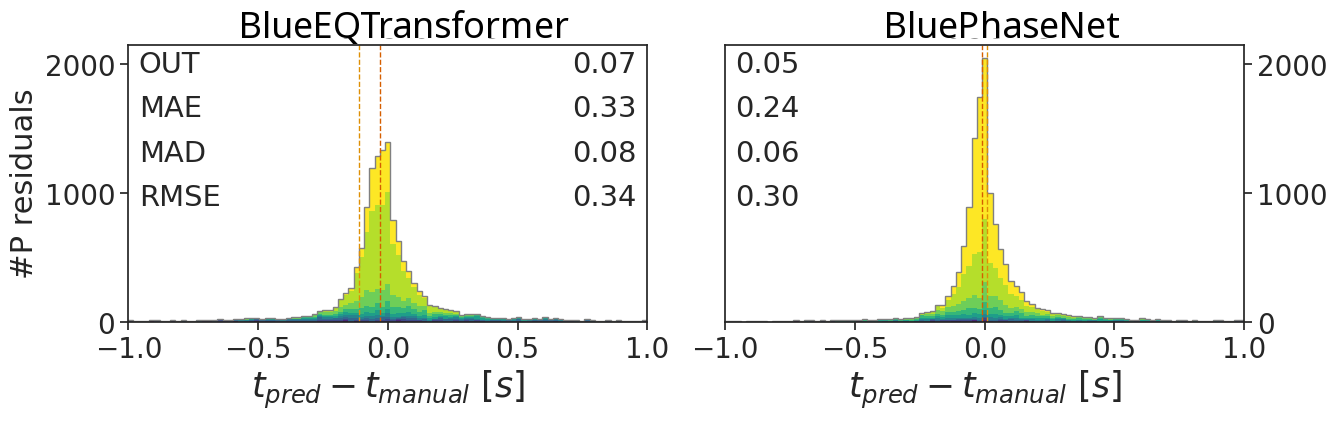}
                \put(10,11){\framebox{\mbox{\LARGE{g}}}}
                \put(56,10){\framebox{\mbox{\LARGE{h}}}}
       \end{overpic}   
\confcb       
\caption{P residuals with confidence stacks for different three-component models, i.e., pickers do not use the hydrophone channel. (a-d) are for pickers trained with onshore station data only, specifically using the models downloaded from the Seisbench platform and trained on the INSTANCE (a-b) and STEAD (c-d) datasets, respectively. (e-f)  show the performance of the three-component model trained with the OBS dataset, i.e., only using the three seismometer components  (pre-training with INSTANCE). The full range to $\pm$10 s is shown in Fig.~\ref{fig:baseline-p-out} and the equivalent plots for S residuals are shown in Fig.~\ref{fig:baseline-s} and~\ref{fig:tl-inst-3comps-s}. (g-h) show how the preferred models perform on the same reduced dataset (i.e., with all 4 components present but only those traces with at least one seismometer observation).}
\label{fig:baseline-p}
\end{figure}

In Fig.~\ref{fig:baseline-p}a-d %
we compare the performance of the OBS picker against EQTransformer and PhaseNet models trained on STEAD and INSTANCE (the performance of the preferred 4-component  pickers, BlueEQTransformer and PhaseNet, on the same subset of picks is shown in Fig.~\ref{fig:baseline-p}g-h).
We chose these models as baseline comparison, as they have been identified as the best performing models in the benchmark study of \citet{benchmarkpaper}.
As the land-based models would not be applicable to traces with only hydrophone components, we ignored those traces for the evaluation, i.e, removing about a sixth of the dataset.

The results show a clear ranking among the models.
Best performing are the \NAMEOFPICKER\ models, which use the hydrophone components.
In particular, the outlier fraction doubles for EQTransformer for some configurations when omitting the hydrophone components.
Most interestingly, the width of the central peak of the residual distribution gets considerably wider.
For P waves on EQTransformer, the MAD increases by 37\% for OBS-trained models without hydrophones and increases three- to seven-fold when using the land-based models.
For P waves on PhaseNet, the decrease in performance when omitting the hydrophones is  less drastic; MAD and outlier fractions barely increase. 
However, the MAD for land-based models drastically increases five- to eleven-fold with respect to the models incorporating hydrophone data. 

Results for the S arrivals closely resemble the results from P waves: the \NAMEOFPICKER~models with hydrophones perform considerably better than those not making use of the hydrophone data, even though one might have expected the hydrophone data to make very little difference for S waves.
In addition, training on OBS data substantially improves performance in terms of both the number of outliers and the residuals.
Interestingly, all three-component land-trained PhaseNet models tend to pick slightly too late for both P- (Fig.~\ref{fig:baseline-p-out}) and S waves (Fig.~\ref{fig:baseline-s}) whereas the EQTransformer equivalents tend to pick too early, except for the STEAD pre-trained variant on P onsets.

Looking into the performance of the individual experiments (Figures~\ref{fig:tl-inst-3comps-p-exp},~\ref{fig:tl-inst-3comps-s-exp}) ALBORAN2009 and LOGATCHEV, we find EQTransformer performing remarkably worse than PhaseNet. The outlier plots (Figures~\ref{fig:tl-inst-3comps-p-exp-out},~\ref{fig:tl-inst-3comps-s-exp-out}) underline this impression: Almost all EQTransformer's picks are too early. The difference between those two experiments, in contrast to the others, is caused by the fact that dropping the hydrophone component only leaves data with a single horizontal component.
This suggests that PhaseNet is better prepared to deal with such degraded data.

In conclusion, these results highlight that the hydrophone component is essential for optimally picking OBS data.
Consequently, pickers trained exclusively on data from land-based stations show clearly inferior performance to those trained on OBS data.

\subsection{Access to Data and Models}

\begin{figure}
  \includegraphics[width=\textwidth]{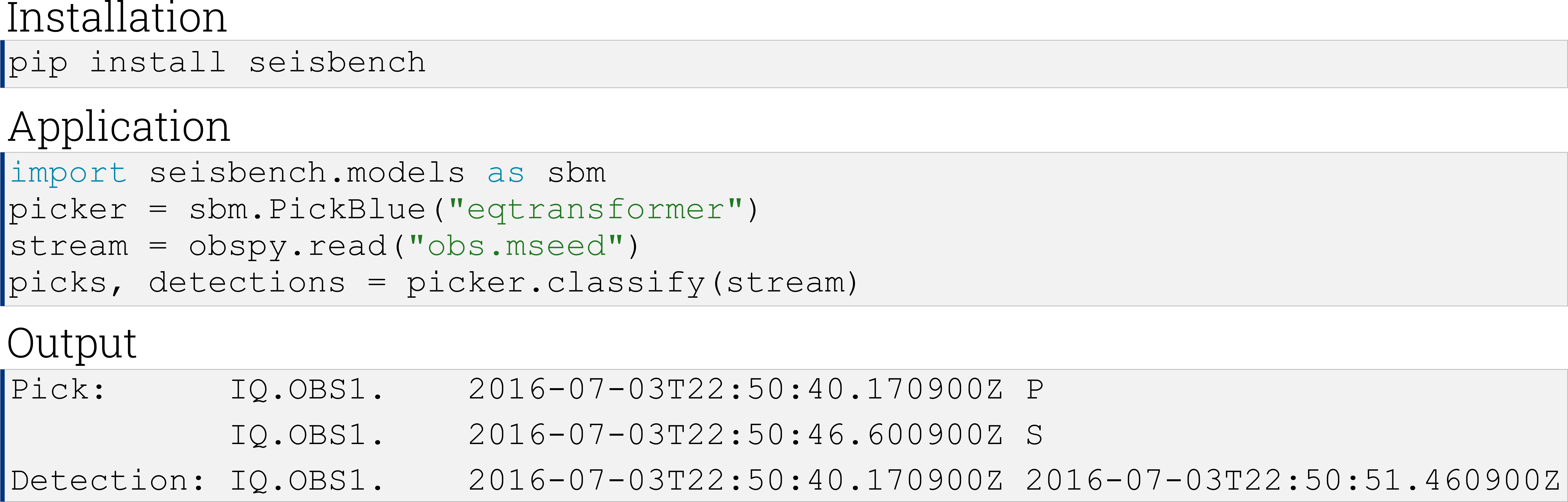}
\caption{Example code for installing and applying {\NAMEOFPICKER} within SeisBench. The lower panel shows the example output.}
\label{fig:ex_application}
\end{figure}

We integrated \NAMEOFPICKER~into the SeisBench package for easy and direct application.
SeisBench is available through PyPI/pip and is licensed under the open GPLv3 license.
\NAMEOFPICKER~can directly be applied to obspy stream objects to obtain phase picks or full characteristic functions.
Fig.~\ref{fig:ex_application} shows an example of how to install and apply \NAMEOFPICKER, as well as an example output.
The implementation automatically applies all necessary preprocessing steps and handles the windowing and reassembling of windows for applying \NAMEOFPICKER~to long input streams.
It makes use of SeisBench's efficient implementation and comes with GPU support and parallelization options to be applied to large-scale datasets.

The OBS reference data set compiled for this study is also available through SeisBench.
It is accessible through the module \textit{seisbench.data} and comes in the standard SeisBench format.
This enables the use of built-in access and filtering methods, as well as the possibility to use the dataset with the SeisBench data generation pipelines.
We hope this access will stimulate the future development of ML models for OBS data for picking but also for other tasks, such as source parameter estimation \cite{munchmeyer2021earthquake}.

\section{Conclusion}

In this study, we trained existing deep learning models for ocean bottom seismometer (OBS) data using four components.
For training and evaluating the models, we compiled a large scale reference data set of labeled OBS waveforms, encompassing data from 15 deployments with more than 150,000 manually labeled phase arrivals of local earthquakes.
Our results show that deep-learning pickers can provide high-quality picks for both P and S phase arrivals.
In particular, for S waves, this is a step-change compared to traditional pickers that often can only pick P phases.

We based our model, \NAMEOFPICKER, on EQTransformer and PhaseNet but added an additional input channel for the hydrophone component.
Overall, the version based on PhaseNet showed slightly better performance than the version based on EQTransformer.
In both cases, the addition of a hydrophone component provides very significant performance improvements.
Even excluding this component, targeted models trained on OBS data substantially improve picking performance compared to models trained exclusively on data from land stations.

Using data from 15 independent deployments allowed us to study performance differences in different settings.
While the number of deployments was insufficient to infer relations between performance and specific deployment conditions, such as tectonic setting or instrument type, we could show that the models expose performance variability between deployments.
This suggests that in application scenarios both picker versions provided by this study should be tested and carefully evaluated.

To allow easy access and application, we make our dataset and models available through the SeisBench library.
We hope that this will foster application to new and existing OBS datasets and will thereby contribute to seismological analysis, such as automated earthquake catalogs.

\section{Acknowledgments}

The authors gratefully thank the Impuls- und Vernetzungsfonds and the Helmholtz Artificial Intelligence Cooperation Unit (HAICU) of the HGF for supporting the REPORT-DL project under the Grant Agreement ZT-I-PF-5-53.
JM acknowledges the support of the Helmholtz Einstein International Berlin Research School in Data Science (HEIBRiDS).
We thank Yu Ren from GEOMAR for providing the manual picks of the BLANCO dataset. 
This work was supported by the Helmholtz Association Initiative and Networking Fund on the HAICORE@KIT partition. 

\section{Data availability}

The continuous BLANCO OBS dataset is archived at the IRIS Data Management System (http://www.iris.edu), network code X9. The waveforms and metadata used in this study, together with the picks, are available through the SeisBench platform (\url{https://github.com/seisbench/seisbench}) and are archived at [DOI will be provided upon final acceptance of the paper]. 

\bibliographystyle{gji}
\bibliography{obs_picking_paper}

\end{document}